\documentclass[twocolumn,showpacs,preprintnumbers,amsmath,amssymb]{revtex4}
\usepackage{graphicx}% Include figure files
\usepackage{dcolumn}% Align table columns on decimal point
\usepackage{bm}% bold math

\begin{document}

\preprint{APS/123-QED}

\title{Field-induced quantum phase in a frustrated zigzag-square lattice}% Force line breaks with \\

%タイトル案
%Observation of thermodynamic properties in a mixed ferromagnetic chain

\author{Hironori Yamaguchi$^{1}$, Kazutoshi Shimamura$^{2}$, Yasuo Yoshida$^{2}$, Akira Matsuo$^{3}$, Koichi Kindo$^{3}$, Kiichi Nakano$^{1}$,\\ Satoshi Morota$^{1}$, Yuko Hosokoshi$^{1}$, Takanori Kida$^{4}$, Yoshiki Iwasaki$^{5}$, Seiya Shimono$^{6}$, Koji Araki$^{7}$, and Masayuki Hagiwara$^{4}$}
%\author{Second Author}
%\email{yamaguchi@p.s.osakafu-u.ac.jp}
\affiliation{
$^1$Department of Physics, Osaka Metropolitan University, Osaka 599-8531, Japan\\
$^2$Department of Physics, Kanazawa University, Ishikawa 920-1192, Japan\\
$^3$Institute for Solid State Physics, the University of Tokyo, Chiba 277-8581, Japan\\
$^4$Center for Advanced High Magnetic Field Science (AHMF), Graduate School of Science, Osaka University, Osaka 560-0043, Japan\\
$^5$Department of Physics, College of Humanities and Sciences, Nihon University, Tokyo 156-8550, Japan\\
$^6$Department of Materials Science and Engineering, National Defense Academy, Kanagawa 239-8686, Japan\\
$^7$Department of Applied Physics, National Defense Academy, Kanagawa 239-8686, Japan
}

%\affiliation{
Second institution and/or address\\
This line break forced% with \\

\date{\today}% It is always \today, today,

\begin{abstract}
This study presents the experimental realization of a spin-1/2 zigzag-square lattice in a verdazyl-based complex, namely ($m$-Py-V-2,6-F$_2$)$[$Cu(hfac)$_2]$.
Molecular orbital calculations suggest the presence of five types of frustrated exchange couplings.
Our observations reveal an incremental increase in the magnetization curve beyond a critical field, signifying a phase transition from the antiferromagnetic ordered state to a quantum state characterized by a 1/2 plateau.
This intriguing behavior arises from the effective stabilization of a zigzag chain by the external fields. 
These results provide evidence for field-induced dimensional reduction in a zigzag-square lattice attributed to the effects of frustration.
\end{abstract}


\maketitle 
%フラストレーション系 ⇒　dimensional reduction 
Frustration has been a pivotal concept in comprehending quantum phenomena in condensed-matter physics, particularly in the 21st century.
In systems with frustration, neighboring spins engage in competing exchange interactions that cannot be simultaneously satisfied, leading to extraordinary quantum phenomena.
An exemplary case is the quantum spin liquid state proposed for a two-dimensional (2D) triangular lattice with spin-1/2 particles~\cite{SL0}, which has sparked extensive and ongoing research~\cite{SL1,SL2}.
Moreover, the attention on quantum spin liquids induced by frustration has been rekindled by the remarkable quantum state anticipated in the Kitaev model of a 2D honeycomb lattice~\cite{kita0,kita1,kita2,kita3}.
Another captivating manifestation of the quantum effects in frustrated 2D lattices is dimensional reduction. 
The exchange couplings within the 2D lattices are partially decoupled to diminish competition and minimize the ground-state energy, thereby stabilizing one-dimensional (1D) states. 
Several frustrated triangular and square lattices have been reported to exhibit 1D quantum behavior stemming from this dimensional reduction~\cite{tri1,tri2,tri3, PF6, SbF6, Zn_gap}.

%zigzag chain の紹介、スピンネマティックの予想　⇒　本系と対応するF-AF with NNNFに言及、近接で実効的ハルデンである点が違い（魅力）
A zigzag chain represents an intriguing 1D spin-lattice with the interplay of low dimensionality and frustration.
Among various types of zigzag chains, the spin-1/2 zigzag chain with competing ferromagnetic (F) nearest-neighbor and antiferromagnetic (AF) next-nearest-neighbor interactions has attracted significant attention in the pursuit of realizing a spin-nematic phase characterized by a specific type of director~\cite{ne1,ne2,ne3,ne4,ne5,ne6,ne_ex1,ne_ex2,ne_ex3,ne_ex4}.
Additionally, numerical simulations have demonstrated that a spin-1/2 zigzag chain composed solely of AF interactions exhibits a diverse range of novel quantum phases under the influence of magnetic fields~\cite{zigAF1,zigAF2,zigAF3,zigAF4}.
In the specific context of the spin-lattice investigated in this study, we assume an alternation of F-AF interactions in the nearest-neighbor coupling.
This F-AF alternating chain exhibits a ground state with distinct topological properties, equivalent to a gapped Haldane state described by a valence bond picture~\cite{haldane}.
On the other hand, the nearest-neighbor uniform chains, whether F or AF, result in a gapless ground state.
Theoretical studies on spin-1/2 zigzag chains composed of F-AF alternating chains with F next-nearest-neighbor interactions have revealed rich quantum phase diagram characterized by the entanglement spectra, which depend on the strength of the frustration~\cite{zigFAF1,zigFAF2,zigFAF3}.

%The spin-1/2 zigzag chain composed only of AF interactions is numerically demonstrated to exhibit varieties of novel quantum phases in magnetic fields.
%The spin model focused on in this study is one example of such a spin-1/2 frustrated square lattice, where three F interactions and one AF interaction form a unit square, as shown in Fig. 1(a). 

%要約
In this Letter, we successfully synthesized single crystals of the verdazyl-based complex  ($m$-Py-V-2,6-F$_2$)$[$Cu(hfac)$_2]$ ($m$-Py-V-2,6-F$_2$ = 3-(3-pyridinyl)-1-(2,6-difluorophenyl)-5-phenylverdazyl, hfac = hexafluoroacetylacetonate). 
Through molecular orbital (MO) calculations, five types of frustrated exchange couplings are predicted. 
Notably, we observe a gradual increase in the magnetization curve beyond a critical field, which serves as evidence for a phase transition from an AF ordered state to a quantum state characterized by a 1/2 plateau.
By applying magnetic fields, the spins on the Cu atoms gradually orient themselves towards the field direction, thus stabilizing an effective zigzag chain composed of radical spins. 
Moreover, the interplay between frustration and quantum fluctuations within this effective zigzag chain disrupts the conventional ordered state, demonstrating the emergence of a field-induced quantum phase in a zigzag-square lattice.

%実験方法
We synthesized ($m$-Py-V-2,6-F$_2$)$[$Cu(hfac)$_2]$ by initially preparing $m$-Py-V-2,6-F$_2$ using the conventional procedure~\cite{gosei}.
The subsequent synthesis of ($m$-Py-V-2,6-F$_2$)$[$Cu(hfac)$_2]$ was accomplished following a previously reported procedure for verdazyl-based complexes~\cite{Zn_alt, Mn_alt, morota,tsukiyama}.
Dark-green crystals of ($m$-Py-V-2,6-F$_2$)$[$Cu(hfac)$_2]$ were obtained through recrystallization from a mixed solvent of CH$_2$Cl$_2$ and $n$-heptan.
Single crystal X-ray diffraction was performed using a Rigaku XtaLAB Synergy-S instrument. 
To measure the magnetic susceptibility, we utilized a commercial SQUID magnetometer (MPMS, Quantum Design) in conjunction with a handmade $^3$He refrigerator, enabling measurements down to 0.59 K~\cite{3He0,3He}. 
The experimental results were corrected by considering the diamagnetic contributions calculated using Pascal's method. 
For specific heat measurements, a commercial calorimeter (PPMS, Quantum Design) employing a thermal relaxation method was employed.
High-field magnetization measurements were conducted using a non-destructive pulse magnet under pulsed magnetic fields.
All experiments were performed using small, randomly oriented single crystals with some polycrystalline samples.

\begin{figure*}[t]
\begin{center}
\includegraphics[width=40pc]{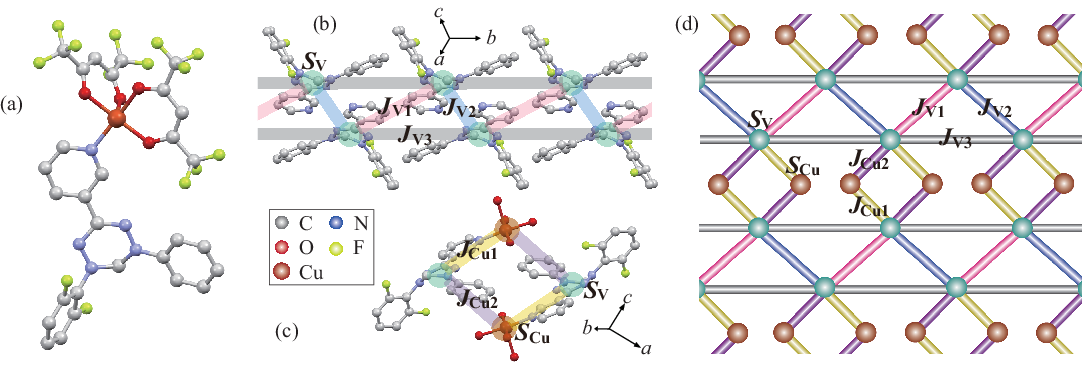}
\caption{(color online) (a) Molecular structure of ($m$-Py-V-2,6-F$_2$)$[$Cu(hfac)$_2]$. Hydrogen atoms are omitted for clarity. Crystal structure of ($m$-Py-V-2,6-F$_2$)$[$Cu(hfac)$_2]$ forming (b) the zigzag chain along the $b$-axis and (c) the square unit. The blue and brown nodes represent the spin-1/2 of the radicals and Cu atoms, respectively. 
The thick lines represent the exchange interactions. (d) Spin-1/2 zigzag-square lattice composed of $J_{\rm{V1}}$, $J_{\rm{V2}}$, $J_{\rm{V3}}$, $J_{\rm{Cu1}}$, and $J_{\rm{Cu2}}$.
}
\end{center}
\end{figure*}

%結晶構造
The molecular structure of ($m$-Py-V-2,6-F$_2$)$[$Cu(hfac)$_2]$ is depicted in Fig. 1(a), where the Cu atom is coordinated by a radical, resulting in a 5-coordinate environment~\cite{supple1}.
The spin of $m$-Py-V-2,6-F$_2$ and Cu$^{2+}$ is 1/2.
The crystallographic parameters at 100 K are as follows: triclinic, space group $P{\bar{\rm{1}}}$, $a$ =  9.6769(4)  $\rm{\AA}$, $b$ = 9.9412(4) $\rm{\AA}$, $c$ = 16.7749(8) $\rm{\AA}$, $\alpha$ = 103.868(4)$^{\circ}$, $\beta$ = 91.103(4)$^{\circ}$, $\gamma$ = 90.358(4)$^{\circ}$, $V$ = 1566.32(12)  $\rm{\AA}^3$, $Z$ = 2, $R$ = 0.0539, and $R_{\rm{w}}$ = 0.1469. 
MO calculations revealed three predominant intermolecular interactions between the radicals, as shown in Fig. 1(b)~\cite{MOcal}.
These interactions are quantified as $J_{\rm{V1}}/k_{\rm{B}}$ = $3.4$ K, $J_{\rm{V2}}/k_{\rm{B}}$ = $-1.7$ K, and $J_{\rm{V3}}/k_{\rm{B}}$ = $-1.2$ K, which are defined within the Heisenberg spin Hamiltonian, given by $\mathcal {H} = J_{n}{\sum^{}_{<i,j>}}\textbf{{\textit S}}_{i}{\cdot}\textbf{{\textit S}}_{j}$.
The resulting spin-lattice corresponds to a spin-1/2 zigzag chain composed of a F-AF alternating chain, with F next-nearest-neighbor interactions.
Additionally, we found that there is not only intramolecular coupling but also a close contact between the radical and copper atom~\cite{supple1}, forming a square unit, as depicted in Fig. 1(c).
The MO calculation indicated that the intramolecular and intermolecular exchange interactions, $J_{\rm{Cu1}}$ and $J_{\rm{Cu2}}$, are F and AF, respectively.
Since the MO estimates tend to overestimate the interactions between verdazyl radicals and transition metals~\cite{morota,tsukiyama}, it is difficult to evaluate their absolute values.
Consequently, assuming all expected interactions, the zigzag chains formed by $J_{\rm{V1}}$, $J_{\rm{V2}}$, and $J_{\rm{V3}}$ are coupled via a square unit formed by $J_{\rm{Cu1}}$ and $J_{\rm{Cu2}}$, resulting in a spin-1/2 zigzag-square lattice, as illustrated in Fig. 1(d).

%The magnitudes of these couplings are evaluated as $J_{\rm{Cu1}}/k_{\rm{B}}$ = $-19$ K and $J_{\rm{Cu2}}/k_{\rm{B}}$ = $4.3$ K.
%Notably, the MO estimates tend to overestimate the intramolecular interactions between verdazyl radicals and transition metals~\cite{morota,tsukiyama}, suggesting that the actual value of $|$$J_{\rm{Cu1}}$$|$ is expected to be smaller.

%磁化率と比熱
Figure 2(a) shows the temperature dependence of the magnetic susceptibility ($\chi$ = $M/H$) at 0.1 T.
The $\chi T$ value exhibits a pronounced decrease as the temperature decreases, indicating the development of AF correlations, as shown in Fig. 2(b). 
In the 10-300 K temperature range, the $\chi$ follows the Curie-Weiss law, with an estimated Weiss temperature of $\theta_{\rm{W}}$ = -1.79 (3) K, indicating dominant AF interactions.
Additionally, a shoulder-like behavior is observed in $\chi$ below approximately 2 K.
At this temperature, the temperature derivative of $\chi$ exhibits a discontinuous change at $T_{\rm{N}}\simeq$ 0.9 K, as shown in the inset of Fig. 2(a), which can be attributed to a phase transition to an AF ordered state. 
If we assume a significant difference in the magnitude of exchange interactions, an energy separation occurs due to the difference in temperature region where the correlation becomes dominant, yielding a multistep change in $\chi T$~\cite{tsukiyama,a26Cl2V,b26Cl2V}.
Because the observed $\chi T$ exhibits a monotonic decrease with decreasing temperature down to $T_{\rm{N}}$ in the present system, we can expect that the magnitudes of the exchange interactions are sufficiently comparable, preventing the energy separation and resulting in an AF ordered state composed of both $S_{\rm{V}}$ and $S_{\rm{Cu}}$ spins.  
The specific heat at zero field exhibits a distinct peak at $T_{\rm{N}}$, indicating a phase transition to the AF ordered state, as shown in Fig. 3.
When magnetic fields are applied, the peak signal disappears above 3 T.

%磁化曲線
Figure 4(a) shows the magnetization curve at 1.5 K above $T_{\rm{N}}$, demonstrating paramagnetic behavior that becomes nearly saturated at approximately 15 T.
Based on the isotropic $g$ value of 2.0 for the organic radicals, the saturation value of 2.1 $\mu_{\rm{B}}$/f.u. suggests that the average $g$ value of $S_{\rm{Cu}}$ is approximately 2.2.
The temperature dependence of the magnetization curve in the low-field region is shown in Fig. 4(b).
Notably, a gradual change is observed above $H_{\rm{c}}\simeq$3 T for $T$ ${\textless}$ $T_{\rm{N}}$, indicating the presence of a 1/2 plateau. 
The slight increase in the plateau phase is considered to be dominated by the thermal excitation effect. 
Furthermore, the field derivative of the magnetization curve ($dM/dH$) exhibits an inflection point at $H_{\rm{c}}$, as shown in the inset of Fig. 4(b).
Considering that the phase transition signal in the specific heat disappears above $H_{\rm{c}}$, it can be inferred that a phase transition from the AF ordered state to a quantum state accompanied by the 1/2 plateau occurs at $H_{\rm{c}}$.

%基底状態の考察
Our analysis focused on the ground state of the anticipated spin-1/2 zigzag-square lattice. 
Experimental observations do not indicate any energy separation associated with significant lattice distortion (the differences in the exchange interactions) that could lead to the formation of nonmagnetic singlet dimers in the system. 
%No experimental result showing energy separation attributed to the difference in the exchange interactions, i.e., distortion of the lattice, is observed, and thus nonmagnetic singlet dimer isn't formed in the system. 
Therefore, the ground state at zero field is expected to be an AF ordered state encompassing all spin sites.
%, while the detailed spin structure is not clarified in the present work.
Figure 1(d) highlights the notable difference in coordination numbers between $S_{\rm{V}}$ (6) and $S_{\rm{Cu}}$ (2). 
The lower coordination number of $S_{\rm{Cu}}$ enhances the effect of polarization by the external magnetic fields.
Consequently, in the low-field region, $S_{\rm{Cu}}$ gradually aligns with the field direction as the field increases, eventually reaching an almost fully polarized state at $H_{\rm{c}}$.
The magnetic moment of 1.1 $\mu_{\rm{B}}$/f.u. at $H_{\rm{c}}$ is consistent with the anticipated value for fully polarized $S_{\rm{Cu}}$ with $g$=2.2 along the field direction.
As a result, the predominantly polarized $S_{\rm{Cu}}$ lacks sufficient degrees of freedom to alter the ground state, effectively leading to a field-induced dimensional reduction where the $S_{\rm{V}}$ chain dominates. 
Furthermore, the frustration and quantum fluctuations within the zigzag chain disrupt the conventional ordered state, giving rise to a field-induced quantum phase accompanied by the 1/2 plateau.
Theoretical studies suggest that the corresponding zigzag chain, composed of F-AF alternating chain with F next-nearest-neighbor interactions, exhibits varieties of gapped quantum phases depending on the exchange parameters~\cite{zigFAF1,zigFAF2,zigFAF3}.
If we assume a singlet state formed by $J_{\rm{V1}}$, the plateau phase observed up to 5 T indicates $J_{\rm{V1}}/k_{\rm{B}}$ ${\textgreater}$ 6.7 K.  
Considering that the evaluated Weiss temperature indicates dominant AF correlations, the actual value of $J_{\rm{V1}}$ is expected to be larger than the MO evaluation, which is consistent with the above prediction.
Furthermore, the internal fields attributed to the coupling with the $S_{\rm{Cu}}$ through $J_{\rm{Cu1}}$ and $J_{\rm{Cu2}}$ can modify the effective field on $S_{\rm{V}}$, leading to the modulation of the plateau region.
Since $J_{\rm{Cu1}}$ and $J_{\rm{Cu2}}$ have opposite signs, the internal fields caused by their couplings are considered to cancel each other. 
Although the precise nature of the quantum state in our system remains uncertain based on our current findings, we expect that the gapped quantum state exhibiting the 1/2 plateau can be attributed to exchange interactions forming the effective zigzag chain.
Figure 4(c) presents a valence bond picture of the quantum state, assuming the presence of a Haldane phase as one of the anticipated quantum phases within the corresponding zigzag chain.
In this representation, two spins coupled by the F interaction $J_{\rm{V2}}$ are considered as effective spin-1, while two spin-1/2 particles on different spin-1 sites form a singlet dimer through the AF interaction $J_{\rm{V1}}$.

%Although we are based on the effective model for a qualitative understanding of the ground state in this consideration, it is difficult to determine the exact exchange parameters.

\begin{figure}
\begin{center}
\includegraphics[width=20pc]{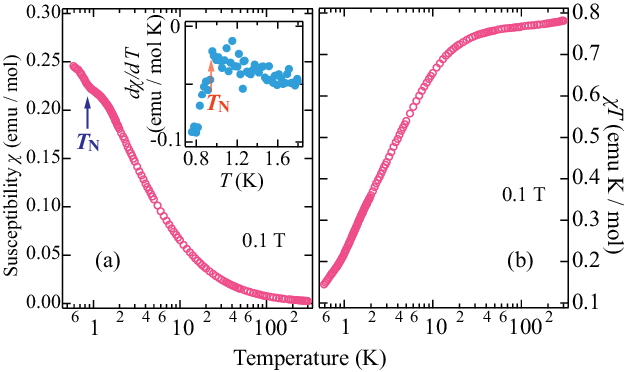}
\caption{(color online) Temperature dependence of (a) magnetic susceptibility ($\chi$ = $M/H$) and (b) $\chi T$ of   ($m$-Py-V-2,6-F$_2$)$[$Cu(hfac)$_2]$ at 0.1 T. 
The inset shows the temperature derivative of $\chi$.
The arrows indicate the phase transition temperature $T_{\rm{N}}$.
}\label{f2}
\end{center}
\end{figure}

\begin{figure}
\begin{center}
\includegraphics[width=16pc]{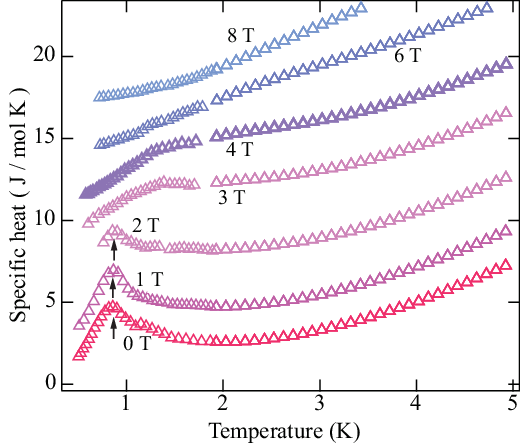}
\caption{(color online) Specific heat of ($m$-Py-V-2,6-F$_2$)$[$Cu(hfac)$_2]$ at various magnetic fields. 
For clarity, the values for 1, 2, 3, 4, 6, and 8 T have been shifted up by 2.0, 5.0, 8.5, 11.0, 14.0 and 17.0 J/ mol K, respectively.
The arrows indicate the phase transition temperature $T_{\rm{N}}$.
}\label{f3}
\end{center}
\end{figure}

\begin{figure}
\begin{center}
\includegraphics[width=20pc]{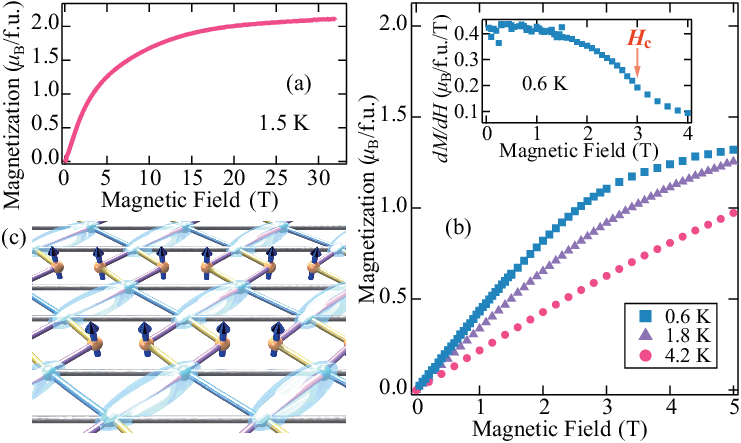}
\caption{(color online) (a) Magnetization curve of ($m$-Py-V-2,6-F$_2$)$[$Cu(hfac)$_2]$ at 1.5 K.
(b) Magnetization curve in the low-field region at various temperatures. 
The inset shows the field derivative of the magnetization curve at 0.6 K, where the arrow indicates the phase transition field $H_{\rm{c}}$.
(c) Valence bond picture of the field-induced quantum phase assuming the Haldane state on the zigzag chain. 
The ovals and arrows represent the valence bond singlet of $S_{\rm{V}}$ via $J_{\rm{V1}}$ and $S_{\rm{Cu}}$ polarized in the external field direction, respectively.
}\label{f4}
\end{center}
\end{figure}

%結論
To summarize, we successfully synthesized single crystals of ($m$-Py-V-2,6-F$_2$)$[$Cu(hfac)$_2]$, a verdazyl-based complex.
MO calculations suggested the presence of three types of intermolecular exchange couplings between the radicals and two types of  exchange couplings between the radical and Cu atom.
The magnetic susceptibility and specific heat indicated a phase transition to an AF ordered state.
The magnetization curve in the ordered phase exhibited a gradual increase above the critical field, indicating a phase transition from the AF ordered state to a quantum state characterized by a 1/2 plateau.
By applying magnetic fields, the spins on the Cu atoms gradually aligned with the field direction, causing the effective zigzag chain composed of radical spins to become prominent.  
This field-induced dimensional reduction destabilized the conventional ordered state, leading to the emergence of a field-induced quantum phase.
We expect that the gapped quantum state featuring the 1/2 plateau is attributed to the effective zigzag chain based on the zigzag-square lattice.
The material studied here provides a platform for investigating a frustrated spin-1/2 zigzag-square lattice with a field-induced quantum phase, which will inspire further numerical studies on its ground state.
It opens up a new research avenue centered on the zigzag-square topology in the field of condensed-matter physics.

This research was partly supported by KAKENHI (Grants No. 23K13065 and No. 23H01127).
A part of this work was performed under the interuniversity cooperative research program of the joint-research program of ISSP, the University of Tokyo.

%%%%%%%%%%%%%%%%%%%%%%%%%%%%%%%%%%%%%%%%%%%%%%%%%%%%%%%%%%%%%%
%%%%%%%%%%%%


\begin{thebibliography}{99}

\bibitem{SL0}
P. W. Anderson, Mater. Res. Bull. \textbf{8}, 153 (1973).

\bibitem{SL1}
L. Balents, Nature (London) \textbf{464}, 199 (2010).

\bibitem{SL2}
L. Savary and L. Balents, Rep. Prog. Phys. \textbf{80}, 016502 (2017).

\bibitem{kita0}
A. Kitaev, Ann. Phys. \textbf{321}, 2 (2006).

\bibitem{kita1}
G. Jackeli and G. Khaliullin, Phys. Rev. Lett. \textbf{102}, 017205 (2009).

\bibitem{kita2}
A. Banerjee, J. Yan, J. Knolle, C. A. Bridges, M. B. Stone, M. D. Lumsden, D. G. Mandrus, D. A. Tennant, R. Moessner, and S.
Nagler, Science \textbf{356}, 1055 (2017).

\bibitem{kita3}
K. Kitagawa, T. Takayama, Y. Matsumoto, A. Kato, R. Takano, Y. Kishimoto, S. Bette, R. Dinnebier, G. Jackeli, and H. Takagi, Nature(London) \textbf{554}, 341 (2018).

\bibitem{tri1}
R. Coldea, D. A. Tennant, A. M. Tsvelik, and Z. Tylczynski, Phys. Rev. Lett. \textbf{86}, 1335 (2001).

\bibitem{tri2}
W. Zheng, J. O. Fjærestad, R. R. P. Singh, R. H. McKenzie, and R. Coldea, Phys. Rev. Lett. \textbf{96}, 057201 (2006).

\bibitem{tri3}
M. Kohno, O. A. Starykh, and L. Balents, Nat. Phys. \textbf{3}, 790 (2007).

\bibitem{PF6}
H. Yamaguchi, Y. Sasaki, T. Okubo, M. Yoshida, T. Kida, M. Hagiwara, Y. Kono, S. Kittaka, T. Sakakibara, M. Takigawa, Y. Iwasaki, and Y. Hosokoshi, Phys. Rev B, \textbf{98}, 094402 (2018)

\bibitem{SbF6}
H. Yamaguchi, Y. Iwasaki, Y. Kono, T. Okubo, S. Miyamoto, Y. Hosokoshi, A. Matsuo, T. Sakakibara, T. Kida, and M. Hagiwara, Phys. Rev B, \textbf{103}, L220407 (2021).

\bibitem{Zn_gap}
H. Yamaguchi, N. Uemoto, T. Okubo, Y. Kono, S. Kittaka, T. Sakakibara, T. Yajima, S. Shimono, Y. Iwasaki, and Y. Hosokoshi, Phys. Rev. B {\bf 104}, L060411 (2021).

\bibitem{ne1} 
N. Shannon, T. Momoi, and P. Sindzingre, Phys. Rev. Lett. \textbf{96}, 027213 (2006).

\bibitem{ne2} 
T. Vekua, A. Honecker, H.-J. Mikeska, and F. Heidrich-Meisner, Phys. Rev. B \textbf{76}, 174420 (2007).

\bibitem{ne3} 
T. Hikihara, L. Kecke, T. Momoi, and A. Furusaki, Phys. Rev. B \textbf{78}, 144404 (2008).

\bibitem{ne4}
M. E. Zhitomirsky and H. Tsunetsugu, Europhys. Lett. \textbf{92}, 37001 (2010).

\bibitem{ne5}
S. Nishimoto, S.-L. Drechsler, R. O. Kuzian, J. van den Brink, J. Richter, W. E. A. Lorenz, Y. Skourski, R. Klingeler, and B. B$\Ddot{\rm{u}}$chner, Phys. Rev. Lett. \textbf{107}, 097201 (2011).

\bibitem{ne6} 
M. Sato, T. Hikihara, and T. Momoi, Phys. Rev. Lett. \textbf{110}, 077206 (2013).


\bibitem{ne_ex1}
M. Mourigal, M. Enderle, B. Fak, R. K. Kremer, J. M. Law, A. Schneidewind, A. Hiess, and A. Prokofiev, Phys. Rev. Lett. \textbf{109}, 027203 (2012).

\bibitem{ne_ex2}
L. E. Svistov, T. Fujita, H. Yamaguchi, S. Kimura, K. Omura, A. Prokofiev, A. I. Smirnov, Z. Honda, and M. Hagiwara, JETP Lett. \textbf{93}, 21 (2011).

\bibitem{ne_ex3}
M. Pregelj, A. Zorko, O. Zaharko, H. Nojiri, H. Berger, L. Chapon, and D. Ar$\check{\rm{c}}$on, Nat. Commun. \textbf{6}, 7255 (2015).

\bibitem{ne_ex4}
M. Pregelj, A. Zorko, M. Gomil$\check{\rm{s}}$ek, M. Klanj$\check{\rm{s}}$ek, O. Zaharko, J. White, H. Luetkens, F. Coomer, T. Ivek, D. G$\acute{\rm{o}}$ngora, H. Berger, and D. Ar$\check{\rm{c}}$on, npj Quantum Mater. \textbf{4}, 22 (2019)


\bibitem{zigAF1} 
K. Okunishi, Y. Hieida, and Y. Akutsu, Phys. Rev. B \textbf{60}, R6953 (1999).

\bibitem{zigAF2} 
N. Maeshima and K. Okunishi, Phys. Rev. B \textbf{62}, 934 (2000).

\bibitem{zigAF3} 
T. Hikihara, T. Momoi, A. Furusaki, and H. Kawamura, Phys. Rev. B \textbf{81}, 224433 (2010).

\bibitem{zigAF4} 
A. K. Kolezhuk, F. H-. Meisner, S. Greschner, and T. Vekua, Phys. Rev. B \textbf{85}, 064420 (2012).

\bibitem{haldane}
I. Affleck, T. Kennedy, E. H. Lieb, and H. Tasaki, Phys. Rev. Lett. \textbf{59}, 799 (1987).

\bibitem{zigFAF1}
K. Hida, K. Takano, and H. Suzuki, J. Phys. Soc. Jpn. \textbf{82}, 064703 (2013).

\bibitem{zigFAF2}
K. Hida, J. Phys. Soc. Jpn. \textbf{85}, 024705 (2016).

\bibitem{zigFAF3}
S. Sahoo, V. M. L. D. P. Goli, D. Sen, and S. Ramasesha, J. Phys.: Condens. Matter \textbf{26}, 276002 (2014).

\bibitem{gosei}
R. Kuhn, Angew, Chem. \textbf{76}, 691 (1964).

\bibitem{Zn_alt}
H. Yamaguchi, Y. Shinpuku, T. Shimokawa, K. Iwase, T. Ono, Y. Kono, S. Kittaka, T. Sakakibara, and Y. Hosokoshi, Phys. Rev. B \textbf{91}, 085117 (2015).

\bibitem{Mn_alt}
H. Yamaguchi, Y. Shinpuku, Y. Kono, S. Kittaka, T. Sakakibara, M. Hagiwara, T. Kawakami, K. Iwase, T. Ono, and Y. Hosokoshi , Phys. Rev. B \textbf{93}, 115145 (2016).

\bibitem{morota}
H. Yamaguchi, S. C. Furuya, S. Morota, S. Shimono, T. Kawakami, Y. Kusanose, Y. Shimura, K. Nakano, and Y. Hosokoshi, Phys. Rev. B \textbf{106}, L100404 (2022).

\bibitem{tsukiyama}
H. Tsukiyama, S. Morota, S. Shimono, Y. Iwasaki, M. Hagiwara, Y. Hosokoshi, and H. Yamaguchi, Phys. Rev. Mater. \textbf{6}, 094417 (2022).

\bibitem{3He0}
Y. Sato, S. Makiyama, Y. Sakamoto, T. Hasuo, Y. Inagaki, T. Fujiwara, H. S. Suzuki, K. Matsubayashi, Y. Uwatoko, and T. Kawae, Jpn. J. Appl. Phys. \textbf{52}, 106702 (2013).

\bibitem{3He}
K. Shimamura, H. Wajima, H. Makino, S. Abe, Y. Haga, Y. Sato, T. Kawae, and Y. Yoshida, Jpn. J. Appl. Phys. \textbf{61}, 056502 (2022).

\bibitem{supple1}
See Supplemental Material at ?????? for details.

\bibitem{MOcal}
The MO calculations were performed using the UB3LYP method in the Gaussian 09 program package. 
The basis sets is 6-31G between radical spins and 6-31G($d$,$p$) between radical and Cu spins. 
For the estimation of intermolecular magnetic interaction, we applied the previously reported evaluation scheme~\cite{MO_method}. 

\bibitem{MO_method}
M. Shoji, K. Koizumi, Y. Kitagawa, T. Kawakami, S. Yamanaka, M. Okumura, and K. Yamaguchi, Chem. Phys.
Lett. \textbf{432}, 343 (2006).

\bibitem{a26Cl2V} 
H. Yamaguchi, T. Okubo, S. Kittaka, T. Sakakibara, K. Araki, K. Iwase, N. Amaya, T. Ono, and Y. Hosokoshi, Sci. Rep. {\bf 5}, 15327 (2015).

\bibitem{b26Cl2V} 
H. Yamaguchi, T. Okubo, K. Iwase, T. Ono, Y. Kono, S. Kittaka, T. Sakakibara, A. Matsuo, K. Kindo, and Y. Hosokoshi, Phys. Rev. B {\bf 88}, 174410 (2013).



%\bibitem{Zn_hone}
%Y. Kono, T. Okabe, N. Uemoto, Y. Iwasaki, Y. Hosokoshi, S. Kittaka, T. Sakakibara, and H. Yamaguchi, Phys. Rev. B \textbf{101}, 014437 (2020).

%\bibitem{Mn_hone}
%Y. Iwasaki, T. Okabe, N. Uemoto, Y. Kono, Y. Hosokoshi, S. Nakamura, S. Kittaka, T. Sakakibara, M. Hagiwara, T. Kawakami,  and H. Yamaguchi, Phys. Rev. B \textbf{101}, 174412 (2020)





\end{thebibliography}
\end{document}